\documentclass[printer]{tMOP2e}
\usepackage{graphicx}

\begin{document}
\title{Resolution in rotation measurements}
\author{Stephen M Barnett and Roberta Zambrini\\
Department of Physics, University of Strathclyde, Glasgow G0 4NG, United Kingdom}
\received{v1.1 14 Feb 2005}

\jname{Journal of Modern Optics}
\jvol{00}
\jnum{00}
\jmonth{January}
\jyear{2005}

\maketitle



\begin{abstract}
\textbf{Abstract.} The limiting resolution in optical interferometry is set by
the number of photons used, with the functional dependence determined by the
state of light that is prepared. We consider the problem of measuring the
rotation of a beam of light about an optical axis and show how the limiting
resolution depends on the total number of {\em quanta of orbital angular
momentum} carried by the light beam.
\end{abstract}

\section{Introduction}
\label{Sect:1}

It has long been recognised that the ultimate accuracy of optical measurements
is set by the quantum nature of light.  Indeed the desire to approach these
quantum  limits was a strong motivation for the study of nonclassical and
particularly squeezed states of light \cite{LoudonKnight}.  The use of coherent
laser  sources typically provides a limiting resolution that is inversely
proportional  to the square root of the mean number of photons used in the
measurement ($N^{-1/2}$).  This can be improved upon by the use of squeezed
states which enhances  the resolution by the square root of the degree of
squeezing ($N^{-1/2}e^{-r}$). The full quantum limit is reached by complete
control of the photon number and gives a quantum limited resolution that it
inversely proportional to the photon  number ($N^{-1}$) \cite{Giovannetti}.

One of the earliest proposals for the application of squeezed light was to 
improve the sensitivity of optical interferometry \cite{Caves}, which was 
demonstrated very soon after the first successful squeezing experiments
\cite{Xiao, Grangier}.  This was followed by a demonstration of enhanced
sensitivity in a spectroscopic measurement \cite{Polzik}.  More recently, it
has been suggested that squeezed light can be used to enhance the resolution
of measurements of small displacements in optical images, or beam
displacements \cite{Claude}.  An experimental demonstration, based on 
squeezed light prepared in a novel `flipped' mode, followed soon afterwards 
\cite{treps2002}.

The quantum limit for detection of phase shifts can be approached using
a balanced interferometer with equal intensity inputs \cite{Holland}.  It
has also been suggested that the same degree of resolution could be achieved
by means of special beam-splitters that send all of the light though one arm 
of the interferometer so that a two-mode `Schr\"odinger-cat' state is prepared
\cite{Jacobson, Nobu}.  The same $N^{-1}$-limited resolution can be obtained
for beam displacements by use of a pair of specially shaped modes, each
having precisely the same number of photons \cite{Barnett2003}.

In this paper we examine the factors limiting our ability to measure the rotation
of a beam about the optical axis.  We will find that, as with interferometric and 
beam-displacement measurements, the resolution depends on the number
of photons used and can be improved by the use of suitable nonclassical 
states of light.  The resolution also depends, however, on the orbital angular
momentum of the light used to make the observation \cite{allen92, OAMbook}.  
We will find that it is the product of the orbital angular momentum per photon, 
$\hbar\ell$, and the total photon number, $N$, that determines the limiting resolution.  
Hence it is the total number of quanta of orbital angular momentum, $N\ell$, that 
sets the minimum detectable rotation.

After some general considerations, Sect.~\ref{Sect:2}, we present two
different schemes to measure small rotations, Sect.~\ref{Sect:3} and 
Sect.~\ref{Sect:4}. A comparison of the resolution achievable by different
measurements concludes the paper, Sect.~\ref{Sect:5}.

\section{General considerations}
\label{Sect:2}

Let us consider a light beam propagating through an {\em image rotator}, that
is a device that rotates an input image about the optical axis.
It is not necessary to specify the form of the rotator, but elementary examples
include a rotating Dove prism \cite{hecht}, or a pair of stationary 
Dove prisms with a fixed relative orientation. The latter arrangement has recently
been used to detect optical angular momentum at the single-photon
level \cite{Leach}. A further example of a beam rotator is a light beam passing
{\em off-axis} through a rotating glass disc, which induces a tangential 
displacement, or rotation,  of the beam \cite{Jones}
\footnote{It has recently been suggested that the dual phenomenon, i.e.   light
carrying orbital angular momentum exerting a torque on a transparent medium,
should also  exist  \cite{miles.rodney.steve}.}.

In this work we consider a beam with an image, or transverse spatial profile,
$u_I(x,y)$ propagating in the $z$ direction through an image rotator.  The
beam after passing through the rotator has a transverse profile 
\begin{eqnarray}\label{eq:rot}
u_O(x,y)=u_I(x\cos\delta\phi +y \sin\delta\phi,y\cos\delta\phi -x 
\sin\delta\phi),
\end{eqnarray}
where $\delta \phi $  is the azimuthal rotation angle and we fix the $z$ axis
as the rotation axis. In Sect.~\ref{Sect:3} and \ref{Sect:4} we will consider
two different beams $u_I$.

\begin{figure}
\begin{center}
\includegraphics[width=8cm]{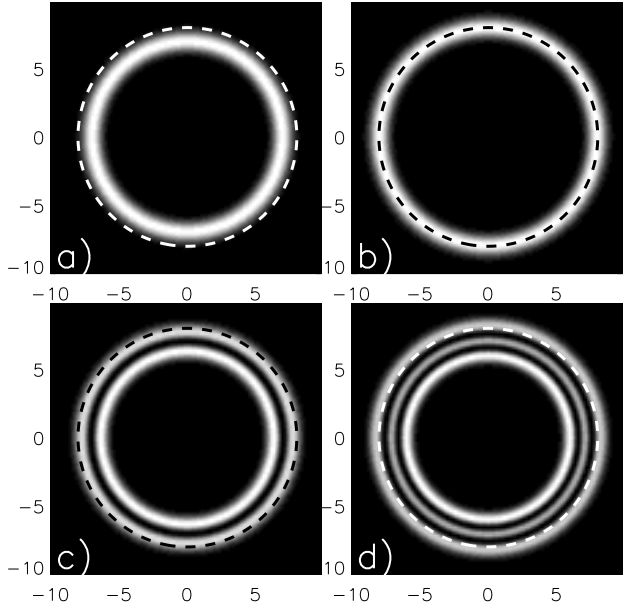}
\end{center}
\caption{\label{fig:2} 
Intensity $\left(\rho^{|l|}e^{-\frac{\rho^2}{2}}L^{|l|}_p(\rho^2)\right)^2$, with
radial coordinate normalized with the beam waist $\rho=r/w_0$. The dashed
circle, with radius $8$ represents the transverse extension of a rotator. Beams
with  $p=0$ have the maximum intensity at $\rho=\sqrt{|l|}$.  a) Intensity for 
$l=49,~p=0$, showing a bright circle with radius $7$. b) For the mode
$l=64,~p=0$ the maximum intensity is at the boundary of the device. For
increasing value of $p$ we observe a spreading in the intensity,  as shown in
c) $l=49,~p=1$ and d)  $l=49,~p=2$.}
\end{figure}

It is natural to describe the beam $u_I$ as superposition of  Laguerre-Gaussian
modes as these are eigenmodes of the $z-$component of angular momentum, which is
the generator of the rotation. This means that the only effect of a rotator on
these modes is to add a constant phase  shift. 
Laguerre-Gaussian  modes, which at  the beam waist have the form
\cite{siegman}:
\begin{equation}\label{eq:normLGmodes}
u_{p\ell}(r,\phi)=\frac{1}{w_0}\sqrt{\frac{p!}{\pi
(|\ell|+p)!}}\exp\left[-\frac{r^2}{2w_0^2}  \right]
\left(\frac{r}{w_0}\right)^{|\ell |}L_p^{|\ell |}\left(\frac{r^2}{w_0^2}\right)
e^{i\ell \phi},
\end{equation}
are labelled
by an angular  index, $\ell$, associated with the angular momentum carried by the
beam \cite{allen92}, and by a radial index, $p$, giving $p+1$ bright rings in the
intensity profile (Fig.~\ref{fig:2}). Modes with  $p=0$ have a single intense ring
with radius \cite{Allen2000}
\begin{eqnarray}\label{eq:r.max}
\bar{r}=w_0\sqrt{|\ell|}.
\end{eqnarray}
Modes with non-vanishing $p$ have a less compact spatial distribution in the
transverse plane (see Fig.~\ref{fig:2}c-d). 

Our study of rotation measurements starts with the realization that the optics used
will, inevitably, have a maximum distance from the optical axis beyond which light
will be lost by the experiment. For simplicity, we suppose that this limit is set
by the radius $R$ of the rotator. This, in turn, sets a maximum value for the
angular momentum that can be carried by a mode propagating through it 
\cite{zambrini2004}. The Laguerre-Gaussian modes with non-zero $p$ extend to a
larger radius than those with the same value of $\ell$  but $p=0$ (see
Fig.~\ref{fig:2}). This means that the largest allowed  angular momentum will occur
for a  $p=0$ mode. For a mode with a bright ring of radius (\ref{eq:r.max}) at the
edges of the device ($\bar{r}=R$), the beam would be strongly diffracted.  
The radial intensity distribution of the Laguerre-Gaussian modes,
for large values of $|\ell|$, has the form 
\begin{eqnarray}\label{eq:LG.radially.Gauss}
|u_{0\ell} (\bar{r}+d)|^2\simeq |u_{0\ell} (\bar{r})|^2 
e^{-d^2/w_0^2},
\end{eqnarray}
so that the intensity tends to be radially distributed like a Gaussian
centred in $\bar{r}$ and with a waist $w_0$. 
Hence we can set the limit for a transmitted Laguerre-Gaussian mode for
\begin{eqnarray}
\bar{r}+w_0=R. 
\end{eqnarray}
From Eq.~(\ref{eq:r.max}) we obtain the maximum angular momentum index 
transmitted by a device with maximum effective radius $R$ as:
\begin{eqnarray}\label{eq:lm}
\ell_{M}=\left(\frac{R}{w_0}-1\right)^2.
\end{eqnarray}

We can use this result to suggest a probable limit for the smallest detectable
rotation $\delta\phi$. Consider the uncertainty relation for rotation angle and
angular momentum \cite{B+P} 
\begin{eqnarray}\label{eq:uncert}
\Delta\phi\Delta L\geq\frac{\hbar}{2}|1-2\pi P(\pi)|,
\end{eqnarray}
where the values of $\phi$ are in the range $[-\pi,\pi]$.
The  form of this uncertainty relation has recently been confirmed
experimentally \cite{sonja}, and states minimizing the  uncertainty product 
(\ref{eq:uncert}) have been derived \cite{pegg}.
For small angular uncertainties we have 
\begin{eqnarray}\label{eq:uncert2}
\Delta\phi\geq\frac{\hbar}{2\Delta L},
\end{eqnarray}
which gives a bound on the minimum possible $\Delta\phi$:
\begin{eqnarray}\label{eq:uncert3}
\Delta\phi\geq\frac{1}{2 \ell_M}.
\end{eqnarray}
For the analogous problem of the optical phase  \cite{P+B} 
the minimum achievable uncertainty is inversely proportional to the mean (or
maximum) photon number ($N$) \cite{summy}. The minimum resolvable phase shift also
seems to be inversely proportional to $N$ \cite{Holland,Barnett2003}.
This suggests that the minimum resolvable rotation given a single photon will be 
\begin{eqnarray}\label{eq:prec}
\delta\phi\propto\ell_M^{-1}.
\end{eqnarray}
We expect that the optimal use of $N$ photons will give a limit 
\begin{eqnarray}\label{eq:prec2}
\delta\phi\propto(N\ell_M)^{-1}.
\end{eqnarray}
The analogy between the uncertainty, $\Delta\phi$, and the resolution,
$\delta\phi$, leads us to refer to (\ref{eq:prec2}) as the `Heisenberg' limit.

\section{Displacement scheme}
\label{Sect:3}

A natural way to measure small angles imparted by an image rotator is through
the displacement of a beam shining the rotator far from the axis, as in Jones
experiment \cite{Jones}. In this scheme the azimuthal displacement gives the
measure of the rotation angle, as shown in Fig.~\ref{fig:1}. Clearly the
resolution is increased by working at the edges of the device, that is at the
maximum distance from the device axis, and with a small size of the light
spot.  In the following we consider a beam with a Gaussian transverse profile,
centred in  $x=r_0,y=0$
\begin{eqnarray}\label{eq:Gauss.in}
u_I(x,y)=\frac{1}{\pi^{1/2}w_0}\exp\left[-\frac{(x-r_0)^2+y^2}{2w_0^2} \right],
\end{eqnarray}
with a small beam waist $w_0$ and large $r_0$, `near' to the edge of the
device. Clearly there are limits for the achievable experimental precision due
simply to the finite size of the  optical elements used. Given a device with a
radial size $R$, than the off-axis Gaussian  (\ref{eq:Gauss.in}) will be
transmitted if $r_0+w_0\sim R$.

The rotated output beam obtained by
Eqs.~(\ref{eq:rot}) and (\ref{eq:Gauss.in})  is
\begin{eqnarray}\label{eq:Gauss.out}
u_O(x,y)=\frac{1}{\pi^{1/2}w_0}
\exp\left[-\frac{(x-r_0\cos\delta\phi)^2+(y-r_0\sin\delta\phi)^2}{2w_0^2} \right].
\end{eqnarray}
The effect of the rotation is to displace the output beam by $\Delta
x=\left[r_0^2(\cos\delta\phi-1)^2+r_0^2\sin^2\delta\phi\right]^{1/2} $.
For small $\delta\phi$ we find
\begin{eqnarray}\label{eq:displ}
\delta\phi=\frac{\Delta x}{r_0},
\end{eqnarray}
so that the resolution achieved measuring small angles in this
scheme depends on the lateral beam position $r_0$ and on the 
precise measurement of the displacement $\Delta x$ between the input and the
rotated light spots.

\begin{figure}
\begin{center}
\end{center}
\caption{\label{fig:1} Scheme based on displacement measurement (picture NA).}
\end{figure}

Small displacements $\Delta x$ are measured with high resolution by shining a
split detector and taking the difference of the light intensities on the two
halves \cite{Claude}. For a perfectly aligned beam the signal detected is
zero, while any small misalignment gives an imbalance in the intensities.
 Given a Gaussian mode in a coherent state with mean photon  number equal to
$N$, the minimum displacement measurable is
\begin{eqnarray}\label{eq:1coh}
\Delta x=\frac{\sqrt{\pi}w_0}{2}\frac{1}{ \sqrt{N}}.
\end{eqnarray}
The standard quantum limit (\ref{eq:1coh}) can be beaten by engineering the spatial
mode impinging on the detector and its statistics.  In particular the input beam is
prepared by superposing an even Gaussian mode (\ref{eq:Gauss.in}) with an odd $
flipped$  mode  $u_I^{odd}(x,y)=u_I(x,y) {\rm sign}(y)$. We note that a flipped
mode is not stable under propagation as it has a discontinuity in $y=0$ that would
be smoothed by diffraction. Nevertheless, it was experimentally possible to beat the
shot noise limit in displacement measurements by shaping this kind of beam
\cite{treps2002}.

In general we have \cite{Barnett2003} 
\begin{eqnarray}\label{eq:1numb}
\Delta x=\frac{\sqrt{\pi}w_0}{2}f(N)
\end{eqnarray}
with $f(N)$ depending on the state in which the modes $u_I^{odd}$ and $u_I$ are
prepared. If the Gaussian mode is in a coherent state with average intensity
$N$ and  the flipped mode is in vacuum then $f(N)=N^{-1/2}$, as in
Eq.~(\ref{eq:1coh}). This is the limit resolution obtained with classical
states, i.e. the standard quantum limit. Better resolution can be achieved if
the flipped mode is prepared in a strongly squeezed state, leading to
$f(N)=N^{-3/4}$.  The best resolution is obtained with highly non-classical
states, for instance by preparing the two modes in number states $|N/2\rangle$. In this
case $f\sim N^{-1}$ and the displacement  $\Delta x \sim N^{-1}$ is the
`Heisenberg limit' mentioned in the previous Section.

From these results for displacement measurements we obtain the
maximum angle resolution of the scheme in Fig.~\ref{fig:1}:
\begin{eqnarray}
\label{eq:dphi} \delta\phi=\frac{ \sqrt{\pi}w_0}{2r_0}f(N).
\end{eqnarray} 
Clearly, $\delta\phi$ depends both on the spatial characteristics of the mode
($w_0$ and lateral displacement $r_0$) and also on the state 
of light (through $f(N)$).   A
decomposition of (\ref{eq:Gauss.in}) in angular momentum eigenmodes allows us to
write  $\delta\phi$ in terms of the angular momentum index $\ell$. In
particular for a Gaussian spot centred far from the axis $z$ ($r_0\gg w_0$) 
there is a large dispersion in the angular momentum spectrum.
We can see this either by writing $u_I(x,y)$ in terms of its angular Fourier
components \cite{vasnetsov} 
\begin{eqnarray}
u_I(x,y)=\frac{1}{\pi^{1/2}w_0}\exp\left(-\frac{x^2+y^2+r_0^2}{2w_0^2}\right)
\sum_{\ell=-\infty}^{\ell=+\infty}I_{|\ell|}
\left(\frac{r_0\sqrt{x^2+y^2}  }{w_0^2}\right)e^{i\ell\phi}
\end{eqnarray} 
or by explicitly constructing its decomposition in
terms of the Laguerre-Gaussian modes
(see Fig.~\ref{fig:histo-gauss}b). The latter procedure is carried out in the
Appendix.
Due to the dispersion in the angular momentum spectrum,
it  is important to consider the constraint, imposed by the extension  $R$
of the rotator, found in Sect.~\ref{Sect:2}. From Eq.~(\ref{eq:lm}) and setting
$r_0+w_0= R$ we find  the maximum resolution in the displacement scheme
\begin{eqnarray}\label{eq:dphi2}
\delta\phi=\frac{\sqrt{\pi}}{2}\frac{1}{\sqrt{\ell_{M}}}f(N). 
\end{eqnarray} 
In Eq.~(\ref{eq:dphi2}) we immediately identify a `geometrical' factor
depending on the angular momentum index and the statistical factor $f$.  In
analogy with the standard quantum limit, obtained by using Gaussian coherent
states in interferometry, we consider the dependence  $\sim{1}/{\sqrt{\ell}}$
in Eq.~(\ref{eq:dphi2}) as the  standard $optical$ limit for rotation
measurements, as it is obtained with Gaussian spatial distributions.  A spatial
Gaussian  mode prepared in a Gaussian coherent state then gives a  combined 
`standard quantum limit' in which the minimum resolvable rotation, $\propto
(N\ell_M)^{-1/2}$, is the inverse of the root square of the number of quanta of
angular momentum. For $r_0 \gg w_0$, the Gaussian mode becomes a good
approximation to an angle-angular momentum minimum uncertainty product state
\cite{pegg} with $<\ell>=0$, $\Delta \ell=r_0/\sqrt{2}w_0=\sqrt{\ell_M/2}$ and 
$\Delta \phi =1/\sqrt{2\ell_M}$. In Fig.~\ref{fig:histo-gauss} the $P(\ell)$
are plotted for  $r_0 =3 w_0$ and $r_0 =10 w_0$ and are compared with Gaussians
having the same variance. The approach to a Gaussian form is an indication of
reaching the minimum uncertainty product limit  \cite{pegg}.

\begin{figure}
\begin{center}
\includegraphics[width=5.5cm]{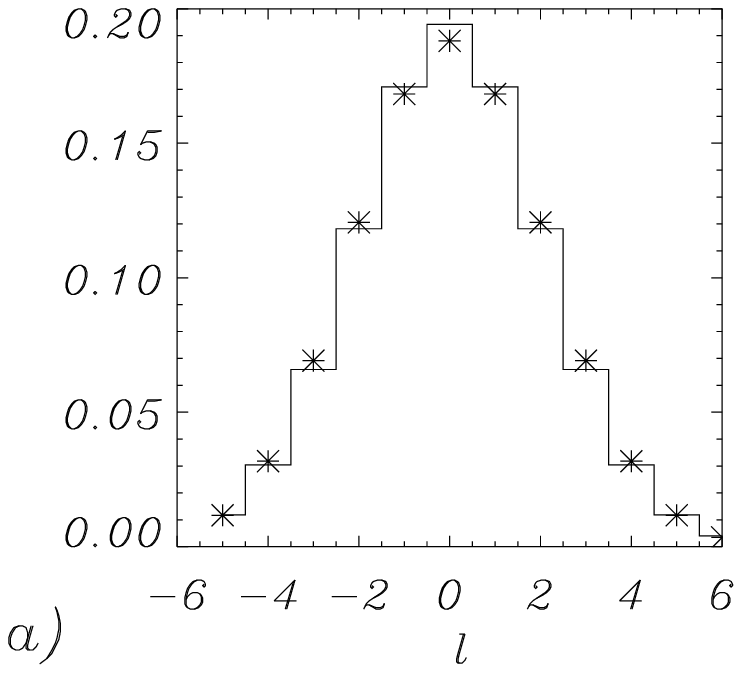}
\includegraphics[width=5.5cm]{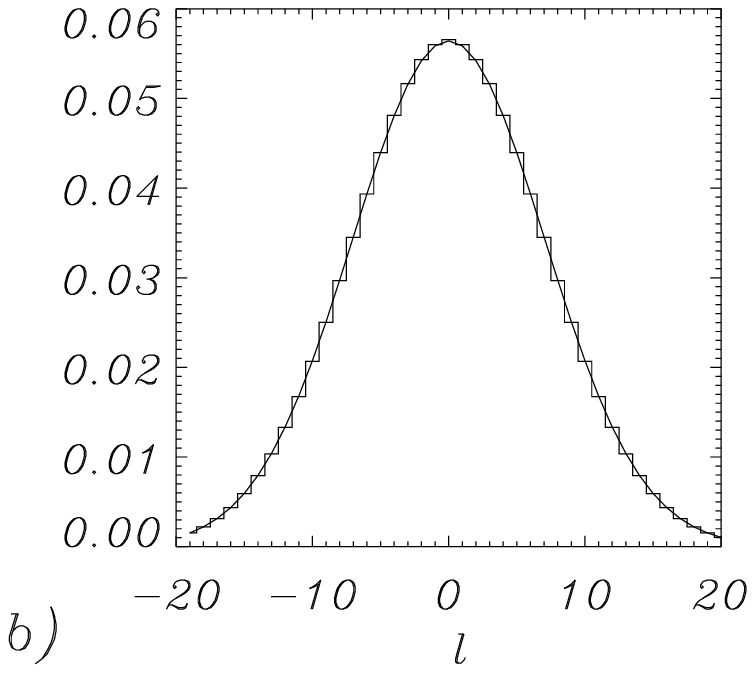}
\end{center}
\caption{\label{fig:histo-gauss} The histograms show the
probabilities $P(\ell)$ given in Eq.~(\ref{eq:Pofell}). The symbols (a) and
smooth line (b) are Gaussians with width 
given by the variance $\Delta \ell = r_0/\sqrt{2}w_0$.
a) $r_0 =3 w_0$. b) $r_0 =10 w_0$.}
\end{figure}

\section{Interferometric scheme}
\label{Sect:4}

If the incoming beam is an angular momentum eigenstate then the only effect of the 
rotator is to add a constant phase shift. 
Interferometers form the basis of phase shift measurements \cite{loudon} and 
so it is natural to consider the interferometer shown in Fig.~\ref{fig:3} to measure
rotations.  The rotator is placed along one of the paths inside the
interferometer.  Here the shift is in the azimuthal spatial profile of the field
and this contrasts with well-known interferometers \cite{interf} designed  to
measure shift in the longitudinal phase of the light beam. 
%
\begin{figure}
\begin{center}
\end{center}
\caption{\label{fig:3} Interferometric phase measurement using 
angular momentum eigenstates.
The single mode annihilation operators are
$\hat a=\int d\vec x v_I(\vec x)\hat a(\vec x)$,
$\hat b=\int d\vec x v_I(\vec x)\hat b(\vec x)$,
where $\hat a(\vec x)$ and $\hat b(\vec x)$ are continuum
annihilation operators \cite{PRA97}. (picture NA)}
\end{figure}
%

Given any mode of the form
\begin{eqnarray}\label{eq:AMeigen.in}
v_I(x,y)=v(r)\exp(i\ell\phi)
\end{eqnarray}
entering in the rotator, the beam at the output will be
\begin{eqnarray}\label{eq:AMeigen.out} 
v_O(x,y)=v_I(x,y)\exp(i\ell\delta\phi).
\end{eqnarray}
We note that the interferometer considered here has recently been used to detect
the angular momentum of single photons \cite{Leach}. In the context of
rotation resolution, we are interested in the smallest  angles
$\delta\phi$ that can be measured with this device.

The rotation through an angle $\delta\phi$ on the beam  (\ref{eq:AMeigen.in})
introduces only a homogeneous phase shift $\ell\delta\phi$ on the whole beam, 
and so it follows that the description  of the interferometer in
Fig.~\ref{fig:3}  -- illuminated by angular momentum eigenmodes -- is
completely equivalent to standard interferometers \cite{interf} measuring
longitudinal phase shifts. We note that to have interference the input modes $
a$ and $ b$ need to have the same angular momentum index ($\ell$).

The difference in the intensities of the two beams emerging  from the
interferometer depends both on the phase shift, here $\ell\delta\phi$, and on
the quantum state of the incoming beams. In particular, when the noise level
has the size of the signal we are at the limit of the  smallest detectable
phase shift
 \begin{eqnarray}\label{eq:dphi4}
\delta\phi=\frac{1}{\ell}f(N),
\end{eqnarray}
with $f(N)= N^{-1/2}, N^{-3/4}, N^{-1}$ depending on the input states of the
modes $ a$ and $b$. We have seen in Sect.~\ref{Sect:2} how the
transverse size of the device sets the limit of the maximum value of $\ell$
of the beam that can be transmitted. By using the maximum allowed angular
momentum we reach the limiting angle resolution  $\propto{1}/{\ell_{M}}$.

It is particularly interesting to consider
the case in which the beams entering in the
interferometer are prepared in the states
$|N/2\rangle|N/2\rangle$ \cite{Holland,Barnett2003}. The angle resolution is then
\begin{eqnarray} \label{eq:limit}
\delta\phi =2.24 \frac{1}{\ell_M N},
\end{eqnarray}
which is the `Heisenberg limit' anticipated in Section \ref{Sect:2}.

\section{Conclusions}
\label{Sect:5}

The resolution attainable in an optical measurement of rotations, $\delta\phi$,
depends on two factors, the number of photons and the orbital angular momentum
content of the  beam. For a displaced Gaussian spot we find,
for a single photon,  that 
$\delta\phi\propto\ell_{M}^{-1/2}$ where $\ell_M$ is the largest angular
momentum index supportable by the image rotator. If the measurement is
performed by using a coherent state with mean photon number  $N$ than we find
that $\delta\phi\propto(N\ell_M)^{-1/2} $, i.e.,  that it is inversely
proportional to the square root of the number of quanta of angular momentum.
Use of nonclassical states of light can enhance the sensitivity by changing the
functional dependence on $N$. In particular, use of correlated number states
can produce a resolution that is proportional to $N^{-1}$. We can also increase
the sensitivity by changing the functional dependence on $\ell_M$. Using
eigenmodes of orbital angular momentum leads to a resolution proportional to  
$\ell_M^{-1}$, with the ultimate `Heisenberg' limit being 
$\propto(N\ell_M)^{-1} $.

We have demonstrated a clear analogy between orbital angular momentum
in rotation measurements and photon number in interferometry.
There are, however, very important practical differences.
Creating states of well defined orbital angular momentum is relatively 
straightforward, while making photon number states is very difficult.
Secondly, enhancement of resolution based on controlling the photon number
requires extremely high efficiencies of photon detection as any losses rapidly
degrade the signal by changing the expected photon number.
Using eigenmodes of  orbital angular momentum, however, is relatively robust as no
matter how many photon are lost, each of the remaining photons still carries 
$\ell\hbar$ units of angular momentum.

\section{Acknowledgements}

This work was supported by the Engineering and Physical Sciences  Research
Council (GR/S03898/01).

\appendices 
\section{Laguerre-Gaussian expansion of a displaced Gaussian beam}

\begin{figure}
\begin{center}
\includegraphics[width=10cm]{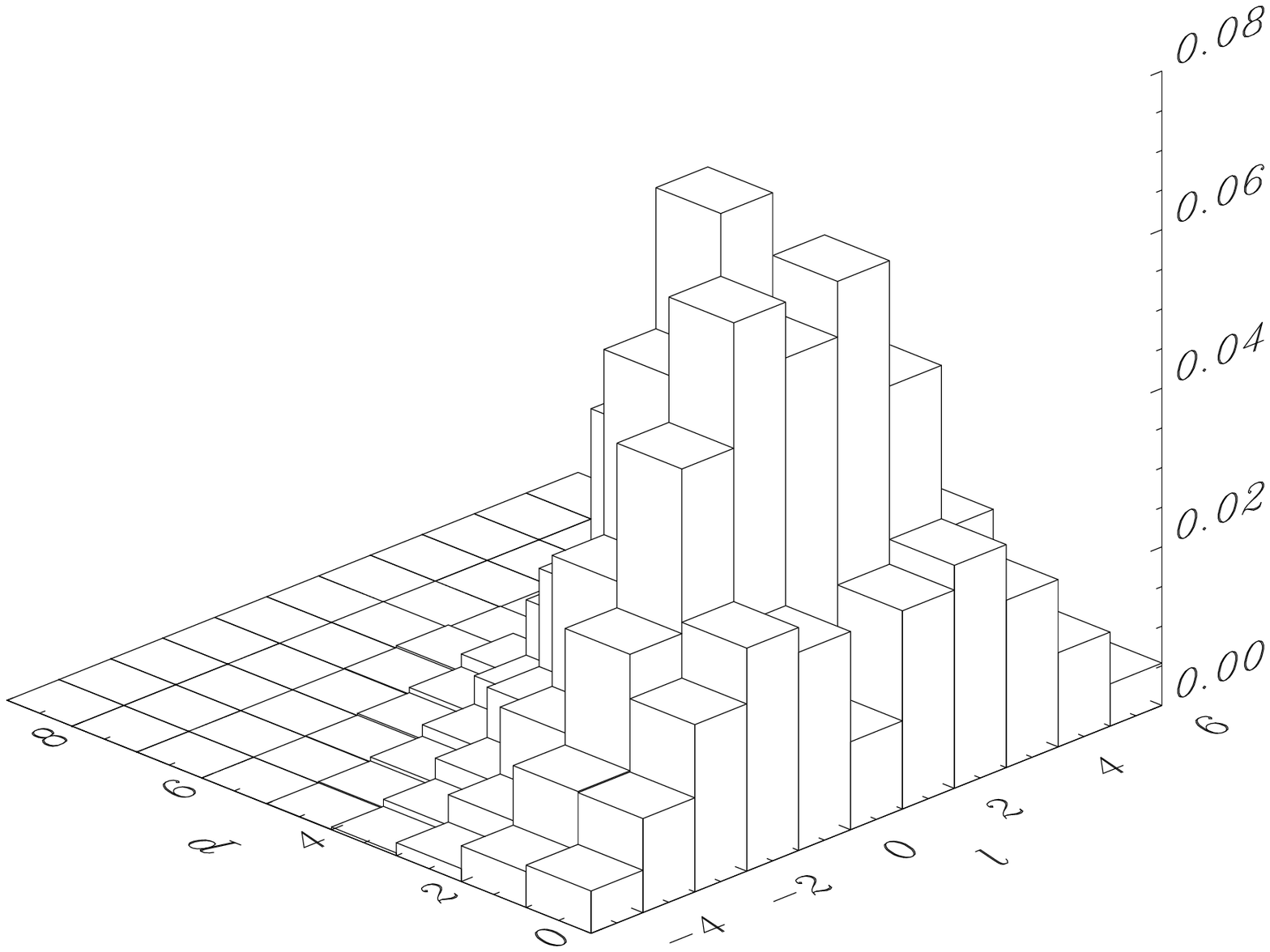}
\includegraphics[width=10cm]{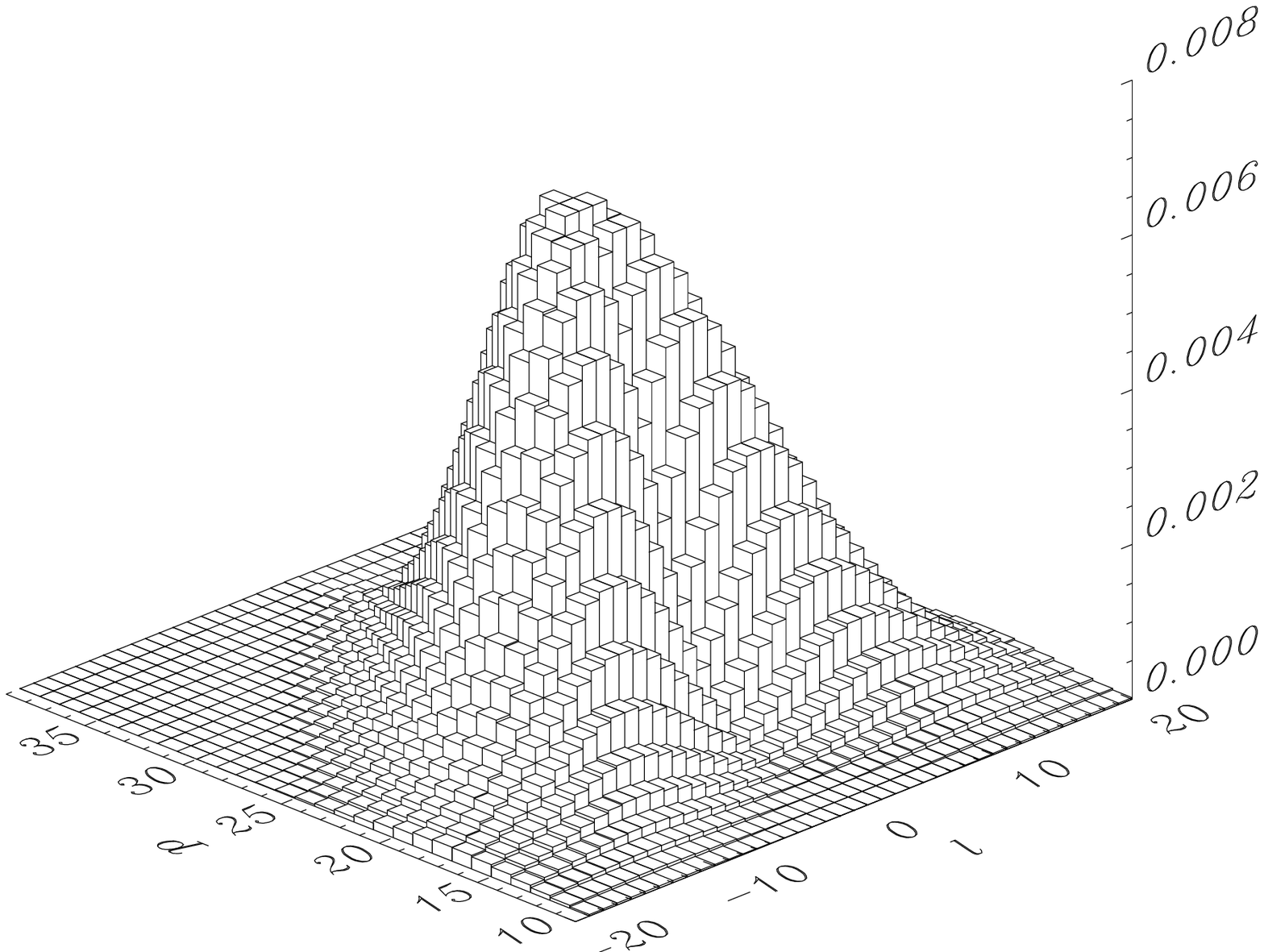}
\end{center}
\caption{\label{fig:histo} Probabilities $|c_{p\ell}|^2=\frac{\exp(r_0^2/2w_0^2)}
{(|\ell|+p)!p!} \left(\frac{r_0}{2w_0} \right)^{4p+2|\ell|}$. Parameter
$r_0/w_0=3$ and $10$.}
\end{figure}

We require the expansion of our displaced Gaussian mode (\ref{eq:Gauss.in})
in terms of the complete set of Laguerre-Gaussian modes: that is we wish
to write (at the beam waist)
\begin{equation}\label{eq:LGexpansion}
u_I(x,y)=\sum_{\ell=0}^\infty \sum_{p=0}^\infty c_{p\ell}u_{p\ell}(r,\phi),
\end{equation}
where $x=r\cos \phi$, $y=r\sin \phi$ and $u_{p\ell}(r,\phi)$ are the normalised 
Laguerre-Gaussian modes (\ref{eq:normLGmodes}).
We find the amplitudes $c_{p\ell}$ by writing both the displaced Gaussian and
the Laguerre-Gaussians as sums  of Hermite-Gaussians and then evaluate their
overlap using the properties of Hermite polynomials.  The displaced Gaussian
can be written in the form
\begin{eqnarray}\label{eq:Gaussexpan}
u_I(x,y)&=&\frac{1}{\pi^{1/2}w_0}\exp\left[-\frac{(x^2 + y^2)}{2w_0^2}  \right]
\exp\left[\frac{r_0 x}{w_0^2}  - \frac{r_0^2}{2w_0^2} \right]  \nonumber \\
&=&\frac{1}{\pi^{1/2}w_0}\exp\left[-\frac{r^2}{2w_0^2} 
\right]\exp\left[-\frac{r_0^2}{4w_0^2}  \right] \sum_{n=0}^\infty
\frac{1}{n!}\left(\frac{r_0}{2w_0}\right)^n H_n\left(\frac{x}{w_0} \right),
\end{eqnarray}
where we have used the generating function for Hermite polynomials
\cite{Abramowitz}.  The Laguerre-Gaussian modes, with $\ell \ge 0$,  can be written 
in the form \cite{Abramochkin}
\begin{eqnarray}\label{eq:LGexpan}
u_{p\ell}(r,\phi)=&&\frac{(-1)^p}{2^{2p+\ell}w_0}\sqrt{\frac{1}{\pi (|\ell|+p)!p!}}\exp\left[-\frac{r^2}{2w_0^2}  \right]
\nonumber \\
&&\times \sum_{k=0}^{\ell + 2p} (2i)^k P_k^{(\ell+p-k,p-k)}(0)H_{\ell + 2p -k}\left(\frac{x}{w_0} \right)H_k\left(\frac{y}{w_0} \right), 
\end{eqnarray}
where 
\begin{equation}\label{eq:Pdef}
P_k^{(n-k,m-k)}(0)=\frac{(-1)^k}{2^kk!}\left. \frac{d^k}{dt^k}[(1-t)^n(1+t)^m]\right|_{t=0}.
\end{equation}
The expansion for negative values of $\ell$ can be obtained by complex conjugation.
We can calculate the coefficients $c_{p\ell}$ using equations (\ref{eq:Gaussexpan}) and 
(\ref{eq:LGexpan}) together with the orthogonality properties of the Hermite 
polynomials:
\begin{eqnarray}\label{eq:cpl}
c_{p\ell}&=&\int_{-\infty}^\infty \int_{-\infty}^\infty  u_{p\ell}^*u_I  dx dy \nonumber \\
&=&(-1)^p\sqrt{\frac{1}{(|\ell|+p)!p!}}\left(\frac{r_0}{2w_0} \right)^{2p+|\ell |}
\exp\left(-\frac{r^2}{4w_0^2}  \right).
\end{eqnarray}
The modulus squared of these amplitudes are plotted in Fig.~\ref{fig:histo},
for $r_0/w_0=3$ and $10$.


It is straightforward to find the fractional power in a displaced Gaussian beam
associated with each value $\ell$, or equivalently the probability, $P(\ell)$, that a single 
photon will be found to have angular momentum $\hbar \ell$:
\begin{equation}\label{eq:Pofell}
P(\ell)=\sum_{p=0}^\infty \left|c_{p\ell}\right|^2 = \exp\left(-\frac{r^2}{2w_0^2}  \right)
I_{|\ell |}\left(\frac{r^2}{2w_0^2}  \right),
\end{equation}
where $I_n$ is the modified Bessel function of order $n$ \cite{Abramowitz}.  That this
probability distribution is normalised follows from the property:
\begin{equation}\label{eq:sumI}
e^z=I_0(z) + \sum_{n=0}^\infty I_n(z).
\end{equation}

\end{document}